\documentclass[a4paper,12pt]{article}
\usepackage[cp1251]{inputenc}
\usepackage{graphics}
\usepackage{amssymb,amsmath}
\usepackage[english]{babel}
\usepackage{indentfirst}

\begin{document}
\bigskip

{\bf Changed Relation Between Radio Flux $F_{10,7}$ And Some Solar
Activity Indices During Cycles 21 - 23}

\bigskip

\centerline {E.A. Bruevich$^{a}$ and G.V. Yakunina$^{b}$ }

\centerline {\it Sternberg Astronomical Institute, Moscow State
 University,}
\centerline {\it Universitetsky pr., 13, Moscow 119992, Russia}\

\centerline {\it e-mail:  {$^a$red-field@yandex.ru,
$^b$yakunina@sai.msu.ru} }\

\bigskip

{\bf Abstract.} A stable cyclicity of correlation coefficients
$K_{corr}$ for some solar activity indices versus 10,7 cm radio flux
$F_{10,7}$ was found. We have analyzed the monthly averages values
of the popular solar indices. These indices are: Wolf numbers,
$F_{10,7}$, 0,1-0,8 nm background, the total solar irradiance, Mg II
UV-index (280 nm core-to-wing ratio) and counts of flares. The
correlation coefficients of the linear regression of these solar
indices versus $F_{10,7}$ were analyzed for every year in solar
cycles 21, 22 and during the unusual cycle 23.
 We found out that the
 correlation coefficients $K_{corr}$ for solar activity
indices versus $F_{10,7}$ (which we determined for each year) show
the cyclic variations with stable period equal to half length of
11-yr cycle (5,5 years approximately).

\bigskip

\bigskip

\vskip12pt
\section{Introduction}
\vskip12pt

We have studied monthly averaged values of six global solar activity
indices during  activity cycles 21, 22 and 23.  Most of these
observed data we used in our paper were published in
Solar-Geophysical Data Bulletin and in Reports of National
Geophysical Data Center Solar and Terrestrial Physics (2012). It's
known that the various activity indices which characterized the
different aspects of the solar magnetic activity correlate quite
well with the most popular solar index such as the sunspot numbers
and with others indices over long time scales. Floyd $\it{et~al.}$
(2005) showed that the mutual relation between sunspot numbers and
also between three solar UV/EUV indices, the $F_{10,7}$ flux and the
Mg II core-to-wing ratio (which is the well known chromospheric UV
index, see (Viereck $\it{et~al.}$ (2001), Viereck $\it{et~al.}$
(2004)), remained stable for 25 years until 2000. At the end of 2001
these mutual relations dramatically changed due to a large
enhancement which took place after actual sunspot maximum of the
cycle 23. The subsequent relative quietness intermediate called the
Gnevyshev gap
 has been the relatively large in the cycle 23.

 All the indices
studied in this paper are very important not only for analysis of
solar radiation which comes from  different altitudes of solar
atmosphere. The most important for solar-terrestrial physics is the
study of solar radiation influence  on the different layers of
terrestrial atmosphere (mainly the solar radiation in EUV/UV
spectral range which effectively heats the thermosphere of the Earth
and so affects to our climate).

 We used in our study the monthly averages
values  for all the activity indices. Such averages allowed us to
take into consideration the fact that the major modulation of solar
indexes are the consequence of
 27-28 days variations of radiative fluxes (these variations correspond
 to the mean solar rotation period).
After monthly averaging we reduced the influence of the rotational
modulation on the observational data.

The unusual cycle 23 was examined carefully: the rise, decline,
minimum and maximum phases were studied separately. This study made
us possible to determine that correlation coefficients of linear
regression for some solar activity indices versus $F_{10,7}$ have
the different values for different cycle's phases.

Vitinsky $\it{et~al.}$ (1986) analyzed solar cycles 18 - 20 and
pointed out that correlation for spot numbers versus radio flux
$F_{10,7}$ does not show the strong linear connection during  all
activity cycle.

Also it was emphasized the importance of statistical study in our
 solar activity processes understanding. To achieve to best agreement in
approximation of spot numbers values by $F_{10,7}$ observations.
Also it was emphasized the importance of statistical study for our
 solar activity processes understanding. To achieve to best agreement in
approximation of spot numbers  with help of $F_{10,7}$ observations
Vitinsky $\it{et~al.}$ (1986) proposed to approximate the dependence
W - $F_{10,7}$ by two linear regressions: the first one - for the
low solar activity (where $F_{10,7}$ less than 150) and the second
one - for the high activity ($F_{10,7}$ more than 150).

In our paper we found out that the linear correlation was violated
not only for maximums of solar activity cycles but for minimums of
the cycles too.

We also analyzed the yearly determined correlation coefficients
$K_{corr}$ for relations of five solar activity indices versus
$F_{10,7}$.

The magnetic activity of the Sun is called the complex of
electromagnetic and hydrodynamic processes in the solar atmosphere
and in the underphotospheric convective zone, see Rozgacheva and
Bruevich (2002) The analysis of active regions (plages and spots in
the photosphere, flocculae in the chromosphere and prominences in
the corona) requires to study the magnetic field of the Sun and the
physics of magnetic activity. This task is of fundamental importance
for astrophysics of the Sun and the stars. Its applied meaning is
connected with the influence of solar active processes on the
Earth's magnetic field.

\section{Global activity indices}

Then we have to say a few words about solar indices studied in
this paper.

The Wolf numbers (W) is a very popular, widely used solar activity
index: the series of W  observations continue more than two hundred
years, see Figure 1.

The solar radio microwave flux at wavelengths 10,7 cm $F_{10,7}$ has
also the longest running series of observations started in 1947 in
Ottawa, Canada and maintained to this day at Penticton site in
British Columbia. This radio emission comes from high part of the
chromosphere and low part of the corona. $F_{10,7}$ radio flux  has
two different sources: thermal bremsstrahlung (due to electrons
radiating when changing direction by being deflected by other
charged participles) and gyro-radiation (due to electrons radiating
when changing direction by gyrating around magnetic fields lines).
These mechanisms give rise to enhanced radiation when the
temperature, density and magnetic fields are enhanced. So $F_{10,7}$
is a good measure of general solar activity. $F_{10,7}$ data are
available at http://radbelts.gsfc.nasa.gov.

The Mg II 280 nm is important solar activity indicator of radiation,
derived from daily solar observations of the core-to-wing ratio of
the Mg II doublet at 279,9 nm provides a good measure of the solar
UV variability and can be used as a reliable proxy to model extreme
UV (EUV) variability during the solar cycle
 Scupin $\it{et~al.}$ (2005). The Mg II observation data were obtained from several satellite's
(NOAA, ENVISAT) instruments. NOAA started in 1978 (during the
$21^{st}$, $22^{nd}$ and the first part of the $23^{rd}$ solar
activity cycles), ENVISAT was launched on 2002 (last part of the
$23^{th}$ solar activity cycle). Comparison of the NOAA and ENVISAT
Mg II index observation data shows that both the MgII indexes agree
to within about 0,5\%. We used both the NOAA and ENVISAT Mg II index
observed data from  Solar-Geophysical Data Bulletin and Scupin
$\it{et~al.}$(2005). Viereck $\it{et~al.}$ (2001) showed an
extremely good fit between the 30,4 nm emission (the main component
of EUV-emission) and the NOAA Mg II index. Viereck $\it{et~al.}$
(2001), Floyd et al. (2005) have noted the linear relationship
between the Mg II index and total solar irradiance.

Solar irradiance is the total amount of solar energy at a given
wavelength received at the top of the earth's atmosphere per unit
time. When integrated over all wavelengths, this quantity is called
the total solar irradiance (TSI) previously known as the solar
constant. Regular monitoring of TSI has been carried out since 1978.
From 1985 the total solar irradiance was observed by Earth Radiation
Budget Satellite (EBRS).

We use the TSI data set from NGDC web site http://www.ngdc.noaa.gov
and combined observational data from National Geophysical Data
Center Solar and Terrestrial Physics (2012). The importance of
UV/EUV influence to TSI variability (Active Sun/Quiet Sun) was
pointed by Krivova \& Solanki  (2008). There were indicated that up
to 63,3 \% of TSI variability is produced at wavelengths below 400
nm. Towards activity maxima the number of sunspots grows
dramatically. But on average the TSI is increased by about 0,1\%
from minimum to maximum of activity cycle. This is due to the
increase amounts of bright features, faculae and network elements on
the solar surface. The total area of the solar surface covered by
such features rises more strongly as the cycle progresses than the
total area of dark sunspots. Some physics-based models have been
developed with using the combined proxies describing sunspot
darkening (sunspot number or areas) and facular brightening (facular
areas, Ca II or Mg II indices), see Frontenla $\it{et~al.}$ (2004),
Krivova $\it{et~al.}$ (2003).

The observations of the X-ray 0,1-0,8 nm background were taken
from Solar Geophysical Data bulletin. There were published the
observational data of The Geostationary Satellite system (GOES),
operated by the United States National Environmental Satellite,
Data, and Information Service (NESDIS). This permanent 0,1-0,8 nm
background monitoring of solar disk at the 0,1-0,8 nm range is a
good indicator of solar corona activity without flares. Several
GOES satellites are still in orbit. The regular background
monitoring of solar disk at the 0,1-0,8 nm range was made with XRS
- X-Ray Irradiance Sensors during GOES 3 - Goes 15 working at the
orbit.

We also analyzed  rapid processes on the Sun - monthly counts of
grouped solar flares (according Solar-Geophysical Data Reports
(2009)) the term 'grouped' means observations of the same event by
different sites were lumped together and counted as one).

\section{Activity indices in the cycle 23}

The recent solar cycle 23 was the outstanding cycle for authentic
observed data from 1849 year. It lasted 12,7 years and was the
longest one for two hundred years of direct solar observations.
This cycle is the second component in the 22-year Hale magnetic
activity cycle but the $23^{rd}$  cycle was the first case of
modern direct observations (from 1849 to 2008 years) when
Gnevyshev-Ol's rule was violated: activity indices in cycle 23 had
their maximum values less then the values in cycle 22 (but
according to Gnevyshev-Ol's rule the cycle 23 must dominate), see
Figure 1.

\begin{figure}[h!]
 \centerline{\includegraphics{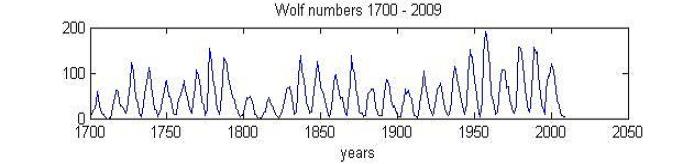}}
\caption{ The time series of annual average of Wolf numbers.
Observations for 1700 to 2005 according to the National Geophysical
Data Center Solar and Terrestrial Physics }\label{Fi:Fig1}
\end{figure}

\begin{figure}[h!]
   \centerline{\hspace*{0.015\textwidth}
               \includegraphics{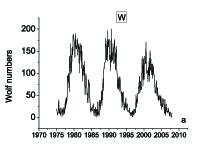}
               \includegraphics{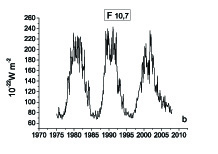}
              }
   \centerline{\hspace*{0.015\textwidth}
               \includegraphics{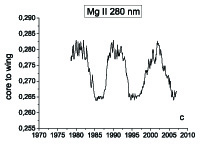}
               \includegraphics{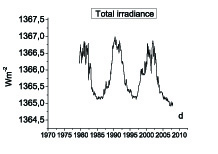}
              }
   \centerline{\hspace*{0.015\textwidth}
               \includegraphics{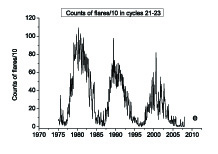}
               \includegraphics{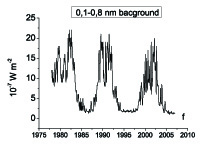}
              }

\caption{The monthly averages for (a) Wolf numbers, (b) $F_{10,7}$
radio flux (low corona), (c) MgII core-to-wing ratio (chromosphere),
(d) TSI (photosphere), (e) counts of flares/10 and (f) 0,1-0,8 nm
background (corona) in cycles 21, 22 and 23.
        }
   \label{F-6panels}
   \end{figure}

Ishkov (2009) pointed that in this unusual cycle 23 the monthly
averages values for Wolf numbers during 8 months exceeded 113 and
most of sunspot groups were less in size, their magnetic fields were
less composite and characterized by the greater lifetime near the
$2^{nd}$ maximum in comparison with values near the $1^{st}$
maximum. In the cycle 23 the $F_{10,7}$ radio flux and the TSI
 have the lowest values from 2007 to 2009 (the
beginning of the cycle 24) from all over of these indices
observation period.

Figures 2 demonstrates that for all activity indices in the
$23^{rd}$ solar activity cycle one can see two maximums separated
one from another on 1,5 year approximately. We see the similar
double-peak structure in the cycle 22 but for the cycle 21 the
double-peak structure is not so evident. We see that there are
displacements in both maximum occurrence time of all these indices
in the $23^{rd}$ solar cycle.

We assume that the probable reason of such double-peak structures is
a modulation of the 11-year fluxes variations by both of the
quasi-biennial and 5,5 year cyclicities of solar magnetic activity,
see Bruevich and Kononovich (2011), Bruevich and Ivanov-Kholodnyj
(2011) . The different time of $1^{st}$ and $2^{nd}$ maximums
appearance may be caused by the difference in fluxes formation
conditions (for our different activity indices) at different
atmosphere's altitudes of the Sun.

Figures 2 also shows that for all solar indices in the cycle 23
 the cavity between two maximums has the relative depth of about
$10-15\%$.

When we have being studied our five activity indices in $23^{rd}$
solar activity cycle we selected three intervals: the rise phase
(from October 1997 to November 1997), the maximum phase (from
November 1997 to July 2002) and the decline phase (from Jul 2002 to
Jan 2006).

The results of correlation study of solar activity indices versus
$F_{10,7}$ in the cycle 23 are presented in Table 1.

Table 1 demonstrates that the maximum values of correlation
coefficients $K_{corr}$ are reached during the rise and decline
phases of the cycles. According to our calculations the highest
values of correlation coefficients $K_{corr}$  we see in connection
between W and $F_{10,7}$. Correlation coefficients $K_{corr}$
between 0,1-0,8 nm background flux versus $F_{10,7}$ are minimal
between all the correlation coefficients determined here. Figure 3
also illustrates the high level of interconnection between solar
activity indices versus $F_{10,7}$ during the unusual cycle 23.

\begin{figure}[h!]
   \centerline{\hspace*{0.015\textwidth}
               \includegraphics{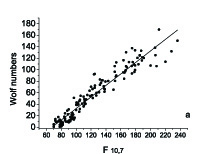}
               \includegraphics{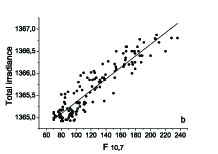}
              }
   \centerline{\hspace*{0.015\textwidth}
               \includegraphics{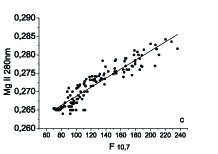}
               \includegraphics{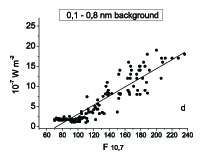}
              }
\caption{Correlations between monthly averages of solar indices and
$F_{10,7}$ radio flux in the cycle 23. (a) Wolf numbers, (b) TSI,
(c) Mg II core to wing ratio and (d) 0,1 - 0,8 nm background.
        }
   \label{F-4panels}
   \end{figure}

The cyclic variation of fluxes in different spectral ranges and
spectral lines at the 11-year time scale
 are widely spread phenomenon for F, G and K stars
 (not only for the case of the Sun).
The chromospheric activity indices (radiative fluxes at the centers
of the H and K emission lines of Ca II - $396,8$ and $393,4$ nm
respectively) for solar-type stars were studied during HK project by
Baliunas {\it at al.} (1995) at Mount Wilson observational program
during 45 years, from 1965 to the present time. Authors of the HK
project supposed that all the solar-type stars with well determined
cyclic activity about 25\% of the time remain in the Maunder minimum
conditions. Some scientists proposed that the solar activity in
future cycle 24 will be very low similar to activity during the
Maunder minimum period. Unlike this Volobuev (2009) predicted the
main parameters of the new $24^{th}$ cycle will be similar to the
 usual activity cycle's parameters (not similar to the cycle's
 characteristics
during the Maunder minimum). At Figure 1 we also see (if we continue
the smoothed curve) the modulation of yearly observations with the
so-called century cyclicity. So we can extrapolate the $24^{th}$
cycle's that Wolf numbers maximum  will be about 150-180.

\begin{table}
\caption{Correlation coefficients between our activity indices and
$F_{10,7}$ at rise, decline and maximum phases of the cycle 23}
\begin{tabular}{ccccc}     
  \hline                   
correlation between     &rise phase        &decline phase                 &  maximum           & all the     \\
  activity indices      & (cycle 23)        &(cycle 23)           &(cycle 23)     &            cycle 23   \\  \hline
  W - $F_{10,7}$          &0.919                 &0.961                 &0.742              &0.939       \\
  MgII- $F_{10,7}$       &0.963                 &0.964                 &0.757              &0.879        \\
  TSI - $F_{10,7}$&0.879               &0.949                 &0.743              &0.920        \\
  0,1-0,8nm - $F_{10,7}$  & 0.899                &0.814                 &0.773              &0.812        \\
  counts fl./10 - $F_{10,7}$& 0.905           &0.895                 &0.785              &0.890        \\
  \hline
\end{tabular}
\end{table}

\section{Changed relation between activity indices and
$F_{10,7}$ in the cycles 21-23}

\begin{figure}[h!]
   \centerline{\hspace*{0.015\textwidth}
               \includegraphics{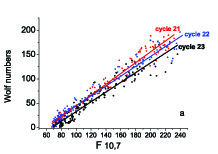}
               \includegraphics{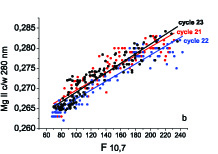}
              }
   \centerline{\hspace*{0.015\textwidth}
               \includegraphics{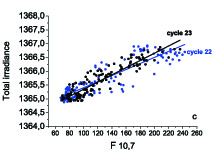}
               \includegraphics{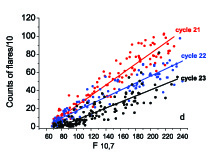}
              }
\caption{Correlation between monthly averages for solar indices and
$F_{10,7}$ radio flux in cycles 21 - 23. (a) Wolf numbers, (b) Mg II
core to wing ratio, (c) TSI and (d) counts of flares/10.}
   \label{F-4panels}
   \end{figure}

We analyzed the interconnection between activity indices W, Mg II
core to wing ratio,TSI and counts of flares versus $F_{10,7}$
during the $21^{st}$, the $22^{nd}$ and the  $23^{th}$ solar cycles.
At Figure 4 we can see correlation between solar indices and radio
flux $F_{10,7}$ in the cycles 21 - 23. The interconnection between
activity indices corresponds to the linear regression equation:

\begin{equation}
    F_{ind}^{cycle}  = a_{ind}^{cycle} + b_{ind}^{cycle} \cdot F_{10,7} \,.
   \end{equation}

were  $F_{ind}^{cycle}$ is our activity index,

$a_{ind}^{cycle}$ is the intercept of linear regression,

$b_{ind}^{cycle}$ is the slope of linear regression.

The results from Table 2 show that coefficients of linear regression
($a_{ind}^{cycle}$ and $b_{ind}^{cycle}$) between solar activity
indices and $F_{10,7}$ differ for different activity cycles (21 -
23). In Table 2 we present standard errors $\sigma$ of the intercept
and the slope values averaged for the cycles 21 - 23. Errors of the
intercept and the slope values $\sigma_a$ and $\sigma_b$ differ
negligible for these cycles. We see that the difference of
regression coefficients of the linear regression for solar activity
indices versus $F_{10,7}$ is the most significant for cycle 23, see
also Figure 4.

Ishkov (2009) pointed that there was very high level of flared
activity in the cycle 21 and very low level of flared activity in
the cycle 23. The Sun's flare activity is an important indicator of
the general level of activity of the atmosphere is also described in
other activity indices, in particular around the solar disk index
and a locally varying flux in the H-alpha, see Bruevich (1995).

We see (Table 2) that the difference of $a_{counts~ fl./10}^{23}$
and $b_{counts~fl./10}^{23}$ values from similar values, determined
for the cycles 21 and 22, is the most significant among all the
different cycle's regression coefficients. Figure 4 also
demonstrates that the flared activity in the $23^{rd}$ cycle almost
twice weaker (counts of flares versus $F_{10,7}$) in comparison to
the flared activity in the $21^{st}$ cycle.

\begin{table}
\caption{Coefficients of linear regression (intercept -
$a_{ind}^{cycle}$ and slope - $b_{ind}^{cycle}$)  between solar
activity indices and $F_{10,7}$.}

\label{T-simple}
\begin{tabular}{ccccc}     
  \hline                   
  indices      & cycle 21        &  cycle 22         &  cycle 23 & errors ($\sigma$)\\ \hline
   $W^{cycle}$         &  a = - 66,4     &  a = -56,43       & a =  -67,27   & $\bar \sigma_a$= 3,0         \\
                           &  b = 1,11         & b = 1.012       & b = 1,001  &   $\bar \sigma_b$= 0,02      \\  \hline
   $F_{MgII}^{cycle}$     & a = 0,257       & a =  0,256      & a = 0,257   &   $ \bar \sigma_a$= 5,25E-04 \\
                           & b = 1,20E-04     & b =  1,09E-04    & b = 1,20E-04  &   $\bar \sigma_b$ = 4,0E-06   \\  \hline
  $F_{TSI}^{cycle}$& -  &a = 1364,312      &   a = 1364,044       &   $\bar\sigma_a$ = 0,058   \\
                           &  -      & b = 0,011     &    b =  0,013   & $\bar \sigma_b$ = 4,4E-04   \\  \hline
$F_{counts fl./10}^{cycle}$& a = -35,38    & a =  -21,33     & a = -25,88    &  $\bar \sigma_a$ = 2,05    \\
                           & b = 0,58      & b =  0,39       & b = 0,33      & $ \bar \sigma_b$ = 0,014  \\
  \hline
\end{tabular}
\end{table}

We have to point out that close interconnection between radiation
fluxes characterized the energy release from different atmosphere's
layers is the widespread phenomenon among the stars of late-type
spectral classes. Bruevich \& Alekseev (2007) confirmed that there
exists the close interconnection between photospheric and coronal
fluxes variations for solar-type stars of F, G, K and M spectral
classes with widely varying activity of their atmospheres. It was
shown that the sum of areas of spots and the values of X-ray fluxes
increase gradually from the Sun and HK project stars with the low
spotted discs to the highly spotted K and M-stars for which Alekseev
\& Gershberg (1996) have constructed the zonal model of the spots
distributed at the star's disks. The variations of activity indices
in the whole 11-yr cycle of the Sun are very similar to the cyclical
variations of the chromospheric fluxes on the stars. So we can
simulate the dependencies which describe the variations of the
indices during the activity cycle for the stars as for the Sun, see
Bruevich and Bruevich (2004).

\begin{figure}[h!]
   \centerline{\hspace*{0.015\textwidth}
               \includegraphics{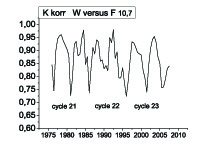}
               \includegraphics{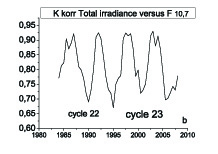}
              }
   \centerline{\hspace*{0.015\textwidth}
               \includegraphics{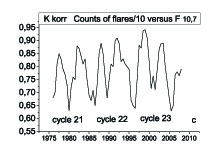}
               \includegraphics{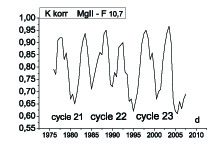}
              }

\caption{Yearly calculated correlation coefficients of linear
regression $K_{corr}$ between  (a) Wolf numbers, (b) TSI, (c) counts
of flares/10 and (d) Mg II UV-index and s $F_{10,7}$
 during solar cycles 21,22 and 23.
        }
   \label{F-4panels}
   \end{figure}

We've calculated yearly averaged values $K_{corr}$ of linear
regression between solar activity indices and $F_{10,7}$ during the
cycles 21, 22 and 23. Yearly averaged values $K_{corr}$ were
determined for each year (we use the monthly averaged values for 12
months of the year with half-monthly time interval, 24 values for
every year).

Figure 5 demonstrates the results of our correlation calculations
between these solar activity indices and $F_{10,7}$  - $K_{corr}$
variations during the cycles 21 - 23. We can see that all the
  $K_{corr}$ values have the maximums at rise and at
decline phases. It's easy to estimate the value of period of
$K_{corr}$ cyclic variations as 5,5 years approximately. We assumed
that this new cyclicity (characterized with period's value equal to
half length of 11-yr  cycle)
 is important for the successful forecasts of the
solar activity indices.

The cyclic behavior of $K_{corr}$ can be explained by following
assumption: we imagine that some activity index flux depends on time
$t$ by the expression:

\begin{equation}
    F_{ind}(t) = F_{ind}^{background}(t) + \Delta F_{ind}^{AR}(t) \,.
   \end{equation}

were $F_{ind}^{background}(t)$ is the background flux which is
increased continuously  with increasing of solar activity and
$\Delta F_{ind}^{AR}(t)$ is the additional flux to the overall flux
from the active regions (AR).

The previous correlation study allows us to consider that
$F_{ind}^{background}(t)$ and $\Delta F_{ind}^{AR}(t)$ are the
linear functions of the background and activity regions levels of
solar activity. In our case we choose the radio flux $F_{10,7}$ as
the best basic indicator of solar activity levels:

\begin{equation}
    F_{ind}^{background}(t) = a_1 + b_1 \cdot
F_{10,7}^{background}(t) \,.
   \end{equation}

\begin{equation}
    \Delta F_{ind}^{AR}(t) = a_2 + b_2 \cdot \Delta
F_{10,7}^{AR}(t) \,.
   \end{equation}

The coefficients  $a_1$ and $b_1$ differ from  $a_2$ and $b_2$. For
W this difference is small, but for 0,1 - 0,8 nm background and
counts of flares the difference between $a_1$, $b_1$ and $a_2$,
$b_2$ is more significant than for W and TSI.

During the rise and decline cycle's phases the dependence $
F_{ind}(t)$ versus $F_{10,7}(t)$ is approximately linear because
coefficients $a$ and $b$ from Table 2 (which describe whole cycle)
are similar to $a_1$ and $b_1$ (see Figure 3 and Figure 4). In this
case we can neglect the relative addition flux from active regions
$\Delta F_{ind}^{AR}(t)$ in respect to $ F_{ind}^{background}(t)$.
So an additional flux from active regions cannot  destroy a balance
in the close linear correlation between $ F_{ind}(t)$ and
$F_{10,7}(t)$ and respective values of $K_{corr}$ reach their
maximum from whole cycle.

During the minimum of activity cycle both values
$F_{ind}^{background}(t)$ and $\Delta F_{ind}^{AR}(t)$ are small,
but the additional flux from active regions can't be neglected in
relation to background flux (which has the minimum values from all
the activity cycle). Therefore values $a$ and $b$ from Table 2 can't
be described to a considerable degree by the linear regression in
cycle's minimum and values of $K_{corr}$ have their minimum values
in the cycle.

During the maximum of activity cycle the values $\Delta
F_{ind}^{AR}(t)$ often exceeds $F_{ind}^{background}(t)$ and
disbalance in linear regression between activity indices increases.
The values of $K_{corr}$ have also their minimum values in the
activity cycle.

\section{Conclusions}

 We have
shown that there are close links between the indices of solar
activity. This can be successfully used for the forecasts of solar
activity on different levels of the solar atmosphere. You need to
take into account that (as we have shown above) such
interconnections are the more powerful in the middle of the cycle
and weaker in moments of maximums and minimums.

In this paper we found out the cyclic behavior of yearly values of
correlation coefficients $K_{corr}$ of linear regression for W, TSI,
Mg II 280 nm and counts of flares versus $F_{10,7}$ during solar
activity cycles 21,22 and 23 (see Figure 5). We showed that yearly
values of $K_{corr}$ have the maximum values at the rise and decline
phases of the activity cycle. Thus the linear connection between
indices is more strong in these cases. It means that the forecasts
of solar indices, based on $F_{10,7}$ observations in our case, will
be more successful during the rise and decline phases of the
activity cycle.

We also determined that the yearly values of $K_{corr}$ are
characterized by cyclic variations with the period that is equal to
half length of 11 - yr main solar cyclicity period (5,5 years
about). Our study of linear regression between solar indices and
$F_{10,7}$ confirms the fact that during the minimum and during the
maximum
 phases of the cycles 21 - 23 the nonlinearity of interconnection between
solar activity indices (characterized the energy release from
different layers of solar atmosphere) increase.

\bigskip

{\bf Acknowledgements} The authors thank the RFBR grant
12-02-00884
 for support of the work.

\end{document}